\documentclass{PoS}
\pdfoutput=1

\usepackage{amsmath}
\usepackage{graphicx}
\usepackage{url}
\usepackage{epsfig}
\usepackage[T1]{fontenc}
\usepackage[latin9]{inputenc}
\usepackage{multirow}

\newcommand{\tb}{\bar t}

\newcommand{\ttbh}{ t \tb H}

\newcommand{\als}{\alpha_{\rm s}}
\newcommand{\shat}{\hat s}

\newcommand{\muf}{\mu_{\rm F}}
\newcommand{\mur}{\mu_{\rm R}}
\newcommand{\mufo}{\mu_{{\rm F},0}}
\newcommand{\muro}{\mu_{{\rm R},0}}
\newcommand{\sigh}{\hat \sigma}

\newcommand{\nn}{\nonumber}

\newcommand{\tosv}{{\scriptscriptstyle \to}}

\def\beq{\begin{equation}}
\def\eeq{\end{equation}}
\def\bear{\begin{eqnarray}}
\def\eear{\end{eqnarray}}

\def\bet34{\beta_{kl}}

\title{	\vspace{-3em}\begin{flushright}\normalsize \sf MS-TP-17-20\\[3em]\end{flushright}
Improving predictions for associated $t\bar{t}H$ production at the LHC: soft gluon resummation through NNLL accuracy}

\ShortTitle{Improving predictions for associated $t\bar{t}H$ production at the LHC: soft gluon resummation through NNLL accuracy}

\author{Anna Kulesza\\
        Institute for Theoretical Physics, WWU M\"unster, D-48149 M\"unster, Germany\\
        E-mail: \email{anna.kulesza@uni-muenster.de}}

\author{Leszek Motyka\\
        Institute of Physics, Jagellonian University, S.\L{}ojasiewicza 11, 30-348 Krak\'ow, Poland\\
        E-mail: \email{leszekm@th.if.uj.edu.pl}}

\author{\speaker{Tomasz Stebel}%
	\\
        Institute of Nuclear Physics PAN, Radzikowskiego 152, 31-342 Krak\'ow, Poland \\
        E-mail: \email{tomasz.stebel@uj.edu.pl}}

\author{Vincent Theeuwes\\
        Department of Physics, SUNY Buffalo, 261 Fronczak Hall, Buffalo, NY 14260-1500, USA\\
        E-mail: \email{vtheeuwe@buffalo.edu}}

\abstract{
In the following we present our recent results on the resummation of soft gluon corrections to the $pp\rightarrow t\bar{t}H$ cross section at the LHC. The resummation was carried out at next-to-next-to-leading-logarithmic (NNLL) accuracy using the Mellin space technique. Obtained results were matched to the NLO cross section. We show that the resummation leads to reduction of scale-variation uncertainty of the total $pp\rightarrow t\bar{t}H$ cross section.

}

\FullConference{The European Physical Society Conference on High Energy Physics\\
                 5-12 July\\
                 Venice, Italy}

\begin{document}

\section{Introduction}

Establishing the properties of the Higgs boson couplings to the Standard Model particles is one of the main tasks of the LHC experiment \cite{deFlorian:2016spz}. The associate production $t\bar{t}H$ offers a direct way to probe the strength of the top--Higgs Yukawa coupling and may be particularly sensitive to physics beyond the Standard Model. Therefore, the improvement of the accuracy for the theoretical predictions is of the central importance.  The next-to-leading-order (NLO) QCD predictions were obtained some time ago~\cite{Beenakker:2001rj, Reina:2001sf}, later they were recalculated and matched to parton showers~\cite{Hirschi:2011pa, Frederix:2011zi, Garzelli:2011vp, Hartanto:2015uka}. Also the QCD-electro weak corrections were calculated~\cite{Yu:2014cka, Frixione:2015zaa}. Finally, the NLO EW and QCD corrections to the hadronic $\ttbh$ production with off-shell top and antitop quarks were obtained~\cite{Denner:2015yca, Denner:2016wet}. The NNLO QCD analysis is currently out of reach so the calculation of soft gluon emission corrections is one of the best way to improve theoretical predictions. In Ref.~\cite{Kulesza:2015vda} we presented the first calculation of the resummed total cross section for the $\ttbh$ production at the next-to-leading-logarithmic (NLL) accuracy. The calculation relied on application of the traditional Mellin-space resummation formalism in the absolute threshold limit, i.e.\ in the limit of the partonic energy $\sqrt{\shat}$ approaching the production threshold $M=2 m_t + m_H$. Subsequently we have performed~\cite{Kulesza:2016vnq} resummation of NLL corrections arising in the limit of $\sqrt{\shat}$ approaching the invariant mass threshold Q, where $Q^2= (p_t +p_{\bar t}+ p_H)^2$. Recently we extended this calculation to the next-to-next-to-leading-logarithmic (NNLL) accuracy \cite{Kulesza:2017ukk}. Threshold resummation can be also performed in the framework of the soft-collinear effective theory (SCET). For the $t\bar{t}H$ process this approach was first applied in Ref.~\cite{Broggio:2015lya} obtaining approximate NNLL and later full NNLL~\cite{Broggio:2016lfj} accuracy.

In this note we report the threshold resummation in the invariant mass limit at the NNLL accuracy using the direct QCD Mellin-space approach ~\cite{Sterman:2013nya}. Taking the Mellin transform allows one to systematically treat the logarithmic terms of the form $\als^n \left[ \log^m (1-\rho )/(1-\rho )\right]_+$, with $m\leq 2n-1$ and $\rho =Q^2/\shat$, appearing in the perturbative expansion of the partonic cross section to all orders in $\als$. In Mellin space these logarithms turn into logarithms of the variable $N$, and the threshold limit $z \to 1$ corresponds to the limit $N \to \infty$.  The Mellin moments of the cross section are taken w.r.t.\ the variable $\rho = Q^2/\shat \ $:
$
\hat\sigma(N,Q^2) = \int_0^1 d \rho \, \rho^{N-1} \hat\sigma( \rho, Q^2 ).
$

We present numerical prediction for the NNLL resummed cross sections matched to the fixed order NLO results. In particular, we study the difference between the NNLL results and the NNLL results with a colour-averaging approximation of the hard function.

\section{Resummation at invariant mass threshold}

\label{s:theory}

The resummed cross section in the Mellin space has the form~\cite{Contopanagos:1996nh}
\bear
\label{eq:res:fact_master}
\frac{d \tilde\sigh^{{\rm (res)}}_{ij\tosv kl B}}{dQ^2}(N,Q^2,\muf^2, \mur^2) 
&=& {\mathrm{Tr}} \left[ \,\mathbf{H}_{ij\tosv kl B} (Q^2,\muf^2, \mur^2) 
 \mathbf{S}_{ij\to klB}(N+1,Q^2,\muf^2, \mur^2) \,  \right] \, \\ \nn
&\times& \Delta^i(N+1,Q^2,\muf^2, \mur^2) \Delta^j(N+1,Q^2,\muf^2, \mur^2),
\eear
where $\mathbf{H}_{ ij\tosv kl B}$ indicates the hard-scattering contributions (including phase space factor), $\mathbf{S}_{ij\to klB}$ contains a soft wide-angle emission corrections and function $\Delta^i(\Delta^j)$ sums the softcollinear and collinear contributions from the incoming parton $i$ (parton $j$) \cite{Catani:1996yz}. The trace in (\ref{eq:res:fact_master}) is taken over colour space.

The soft function is given by a solution of the renormalization group equation~\cite{KS2,Czakon:2009zw}:
\beq
 \mathbf{S}_{ij\to klB}(N,Q^2,\muf^2, \mur^2)= \mathbf{\bar{U}}_{ij\tosv kl B}(N, Q^2,\muf^2, \mur^2 )\  \mathbf{\tilde S}_{ij\to klB}(\als(Q^2/{\bar N^2})) \mathbf{{U}}_{ij\tosv kl B}(N,Q^2,\muf^2, \mur^2), 
\label{eq:soft:evol}
\eeq
where $\mathbf{\tilde S}_{ij\to klB}$ plays a role of a boundary condition.

Both hard function and soft matrix initial condition can be calculated perturbatively \cite{KS2,Dixon:2008gr}: $\mathbf{H}_{ ij\tosv kl B}= \mathbf{H}^{\mathrm{(0)}}_{ij\to klB} + \frac{\als}{\pi}\mathbf{H}^{\mathrm{(1)}}_{ij\to klB} +\ldots$ and $\mathbf{\tilde S}_{ij\to klB}= \mathbf{\tilde S}^{\mathrm{(0)}}_{ij\to klB} + \frac{\als}{\pi}\mathbf{\tilde S}^{\mathrm{(1)}}_{ij\to klB} + \ldots$.
%\bear
%\mathbf{H}_{ ij\tosv kl B}&=& \mathbf{H}^{\mathrm{(0)}}_{ij\to klB} + \frac{\als}{\pi}\mathbf{H}^{\mathrm{(1)}}_{ij\to klB} +\ldots \\
%\mathbf{\tilde S}_{ij\to klB}&=& \mathbf{\tilde S}^{\mathrm{(0)}}_{ij\to klB} + \frac{\als}{\pi}\mathbf{\tilde S}^{\mathrm{(1)}}_{ij\to klB} + \ldots 
%\label{eq:soft:nlo}
%\eear
At the NNLL accuracy knowledge of $\mathbf{\tilde S}^{\mathrm{(1)}}_{ij\to klB}$ and $\mathbf{H}^{\mathrm{(1)}}_{ij\to klB}$ is required \cite{Beenakker:2011sf,Beenakker:2013mva} whereas for NLL only leading terms  $\mathbf{H}^{\mathrm{(0)}}_{ij\to klB}$, $\mathbf{\tilde S}^{\mathrm{(0)}}_{ij\to klB}$ are needed. Hard function $\mathbf{H}_{ ij\tosv kl B}$ carries no dependence on $N$. The dependence on $N$ in the soft function $\mathbf{\tilde S}_R$ enters only through the argument of $\als$ and (after expanding in $\als$) results in $\als^2(\mur^2)\log N$ term.

The soft function evolution matrices  $\mathbf{{U}}_{ij\tosv kl B}$, $\mathbf{\bar{U}}_{ij\tosv kl B}$ contain  logarithmic enhancements due to soft wide-angle emissions~\cite{Kidonakis:1998bk}. $\mathbf{{U}}_{ij\tosv kl B}$ is defined as a path-ordered exponent 
$$\mathbf{U}_{ij\to klB}\left(N, Q^2,\muf^2, \mur^2\right)=\mathrm{P}\exp\left[\int_{\muf}^{Q/\bar{N}}\frac{dq}{q}\pmb{\Gamma}_{ij\to klB}\left(\alpha_{\mathrm{s}}\left(q^{2}\right)\right)\right],$$ where the
%\begin{eqnarray}
%\label{eq:soft:evolfact}
%\mathbf{\bar{U}}_{ij\to klB}\left(N, Q^2,\muf^2, \mur^2\right)  &= & \bar{\mathrm{P}}\exp\left[\int_{\muf}^{Q/\bar{N}}\frac{dq}{q}\pmb{\Gamma}_{ij\to klB}^{\dagger}\left(\alpha_{\mathrm{s}}\left(q^{2}\right)\right)\right], \\
%\mathbf{U}_{ij\to klB}\left(N, Q^2,\muf^2, \mur^2\right)  &= & \mathrm{P}\exp\left[\int_{\muf}^{Q/\bar{N}}\frac{dq}{q}\pmb{\Gamma}_{ij\to klB}\left(\alpha_{\mathrm{s}}\left(q^{2}\right)\right)\right], \nonumber
%\end{eqnarray}
%where $\mathrm{P}$ and  $\bar{\mathrm{P}}$ denote the path- and reverse path-ordering in the variable $q$, respectively. 
soft anomalous dimension is calculated as a perturbative function in $\als$,
$\pmb{\Gamma}_{ij\to klB}\left(\als\right)= \left(\frac{\als}{\pi}\right) \pmb{\Gamma}^{(1)}_{ij\to klB}$ $+\left(\frac{\als}{\pi}\right)^2 \pmb{\Gamma}^{(2)}_{ij\to klB}+\ldots$ \cite{Kulesza:2015vda,Ferroglia:2009ep}.
In order to diagonalize the one-loop soft anomalous dimension matrix we make use of the transformation~\cite{Kidonakis:1998bk}:
\begin{equation}
\pmb{\Gamma}^{(1)}_{R}  =  \mathbf{R}^{-1}\pmb{\Gamma}^{(1)}_{ij\to klB} \mathbf{R} 
\end{equation}
and other matrices are transformed using diagonalization matrix $\mathbf{R}$:
$\pmb{\Gamma}^{(2)}_{R} = \mathbf{R}^{-1}\, \pmb{\Gamma}^{(2)}_{ij\to klB}\, \mathbf{R}, \ $ 
$\mathbf{H}_R  =  \mathbf{R}^{-1}\, \mathbf{H}_{ij\to klB} \, \left(\mathbf{R}^{-1}\right)^{\dagger}, \ $ 
$\mathbf{\tilde S}_{R} =  \mathbf{R}^{\dagger}\, \mathbf{\tilde S}_{ij\to klB}\, \mathbf{R}. $
In the $\mathbf{R}$-representation the evolution factor $\mathbf{U}_R$ (similarly $\mathbf{\bar{U}}_R$) can be written at NNLL accuracy as \cite{Buras:1979yt,Ahrens:2010zv}:
\beq
\label{eq:UR}
\mathbf{U}_R(N,Q^2,Q^2, \mur^2 )=\left(\mathbf{1}+\frac{\alpha_{\mathrm{s}}(\mur^2)}{\pi[1-2 \als(\mur^2) b_0 \log N]}\mathbf{K}\right)
\left[e^{\,g_s(N)\overrightarrow{\lambda}^{(1)}} \right]_{D}
\left(\mathbf{1}-\frac{\alpha_{\mathrm{s}}(\mur^2)}{\pi}\, \mathbf{K}\right),
\eeq
where
$K_{IJ}=\delta_{IJ}{\lambda}^{\left(1\right)}_{I}\frac{b_1}{2b_0^2}-\frac{\left(\pmb{\Gamma}^{(2)}_R\right)_{IJ}}{2\pi b_0+\lambda^{\left(1\right)}_{I}-\lambda^{\left(1\right)}_{J}}\,
$ and $\lambda^{(1)}_I$ are the eigenvalues of $\pmb{\Gamma}^{(1)}_{ij\to klB}$.
By $\left[e^{\,g_s(N)\overrightarrow{\lambda}^{(1)}} \right]_{D}$ we have denoted diagonal matrix with exponentiated eigenvalues on diagonal and $g_s(N)$ is a function which resumms logarithms of $N$ (see \cite{Kulesza:2017ukk} for expression), $b_0$ and $b_1$ are the first two coefficients of expansion $\beta_{\mathrm{QCD}}$ in $\als$. 

The resummation-improved cross sections for the $pp \to \ttbh$ process are
obtained through matching the  resummed expression with 
the full NLO cross sections
\beq
\label{hires}
 \frac{d\sigma^{\rm (matched)}_{h_1 h_2 \tosv klB}}{dQ^2}(Q^2,\muf^2, \mur^2) = 
\frac{d\sigma^{\rm (NLO)}_{h_1 h_2 \tosv kl B}}{dQ^2}(Q^2,\muf^2, \mur^2)
+   \frac{d \sigma^{\rm
  (res-exp)}_
{h_1 h_2 \tosv kl B}}{dQ^2}(Q^2,\muf^2, \mur^2) 
\eeq
 with
\beq
\label{invmel}
 \frac{d \sigma^{\rm
  (res-exp)}_{h_1 h_2 \tosv kl B}}{dQ^2} \! =   \sum_{i,j}\,
\int_{\sf C}\,\frac{dN}{2\pi
  i} \; \rho^{-N} f^{(N+1)} _{i/h{_1}} \, f^{(N+1)} _{j/h_{2}}
 \! \left[ 
\frac{d \tilde\sigh^{\rm (res)}_{ij\tosv kl B}}{dQ^2} 
-  \frac{d \tilde\sigh^{\rm (res)}_{ij\tosv kl B}}{dQ^2}
{ \Bigg. \Bigg|}_{\scriptscriptstyle({\rm NLO})}\, \! \right], 
\eeq
where ${d \tilde\sigh^{\rm (res)}_{ij\tosv kl B}}/{dQ^2}$ is given by (\ref{eq:res:fact_master}) and $ {d \tilde\sigh^{\rm (res)}_{ij\tosv kl B}}/{dQ^2}{ \big. \big|}_{\scriptscriptstyle({\rm NLO})}$ represents its perturbative expansion truncated at NLO. $f^{(N)}_{i/h} $ is a Mellin moment (with respect of $x$ variable) of 
parton distribution function for parton $i$ in hadron $h$.

Apart from the full NNLL cross sections we also consider the NLL results, obtained by taking $\mathbf{H}_{ ij\tosv kl B}= \mathbf{H}^{\mathrm{(0)}}_{ij\to klB}$, $\mathbf{\tilde S}_{ij\to klB} = \mathbf{\tilde S}^{\mathrm{(0)}}_{ij\to klB}$, $\mathbf{K}=\mathbf{0}$ and dropping NNLL terms in $\Delta$ and $g_s$. Additionally, we study the NNLL results where an approximation to the non-logarithmic terms,
forgoing the colour structure of the one-loop hard corrections, has been applied. In this approximation, to which we refer as "NNLL  $\bar {\cal C}$", we
calculate a hard coefficient $\bar {\cal C}^{(1)}$  as a colour average of $\cal O(\als)$ non-logarithmic contributions:
\beq
\bar {\cal C}^{(1)}_{ij\tosv klB}(Q^2,\mu_F^2, \mu_R^2) = \mathrm{Tr}\left[\mathbf{H}^{\mathrm{(1)}}_{R} \mathbf{\tilde S}^{\mathrm{(0)}}_{R}+ \mathbf{H}^{\mathrm{(0)}}_{R} \mathbf{\tilde S}^{\mathrm{(1)}}_{R} \right]/  \mathrm{Tr}\left[\mathbf{H}^{\mathrm{(0)}}_{R} \mathbf{\tilde S}^{\mathrm{(0)}}_{R} \right]
\label{Ccoeff_Def}
\eeq
Because of the form of $\mathbf{\tilde S}^{\mathrm{(0)}}$ \cite{Kulesza:2017ukk}, the one-loop hard coefficient $\bar {\cal C}^{(1)}$ involves only virtual hard contributions summed over colour channels. Accounting for the $\bar {\cal C}^{(1)}$ coefficient, Eq.~(\ref{eq:res:fact_master}) is then transformed
into (we skip arguments for simplicity and write it in $\mathbf{R}$-representation):
\beq
\frac{d\tilde \sigh^{\ (\mathrm{NNLL} \ \bar {\cal C})}_{ij\tosv kl B}}{dQ^2}= \left( 1 + \frac{\als}{\pi} {\bar {\cal C}}^{(1)}_{ij\tosv klB}\right)
{\mathrm{Tr}}\left[ \,\mathbf{H}^{\mathrm{(0)}}_R\  \mathbf{\bar{U}}_R \  \mathbf{\tilde S}^{\mathrm{(0)}}_R \, 
\mathbf{{U}}_R \right] \,\Delta^i \ \Delta^j.
\label{eq:res:fact_diag}
\eeq

\section{Numerical results}

%\begin{figure}[t]
%\centering
%\includegraphics[width=0.45\textwidth]{Htt-scaleQ-PDF4LHC_NLO-Q_14TeV.pdf}
%\includegraphics[width=0.45\textwidth]{Htt-scaleMover2-PDF4LHC_NLO-Q_14TeV.pdf}
%\caption{Comparison between the expansion of the resummed expression Eq.~(\label{eq:res:fact_master}) up to NLO accuracy in $\alpha_{\mathrm{s}}$, the full NLO result and the NLO result without the $qg$ channel contribution as a function of the scale $\mu=\muf=\mur$.} 
%\label{f:NLL_expansion_vs_NLO}
%\end{figure}

In this section we present our numerical results obtained for $\sqrt S=14$ TeV. Results for the total cross section are obtained by integrating out the invariant mass distribution (\ref{hires}) over invariant mass $Q$. We use $m_t=173$ GeV, $m_H=125$ GeV and PDF4LHC15{\_}100 sets~\cite{Butterworth:2015oua} . The NLO cross section is calculated using the aMC@NLO code~\cite{Alwall:2014hca}. For the evaluation of the first-order hard function matrix $\mathbf{H}^{\mathrm{(1)}}_{ij\to klB}$ the one-loop virtual corrections to the process (decomposed into various colour transitions $IJ$) are required. We extract them numerically by modification of the publicly available {\tt PowHel} implementation of the $\ttbh$ process~\cite{Garzelli:2011vp}.

Two choices for the central value of the renormalization and factorization scales are used: $\mu_0=\mufo=\muro=Q$ and $\mu_0=\mufo=\muro=M/2=m_t+m_H/2$. The former choice is motivated by invariant mass $Q$ being the natural scale for the invariant mass kinematics used in resummation. The latter choice of the scale is often made in the NLO calculations, see e.g.~\cite{Beenakker:2001rj}.

%In Figure \ref{f:NLL_expansion_vs_NLO} we show the comparison between: NLO cross section without a contribution from the $qg$ channel (blue dotted), expansion of the NNLL-resummed cross section up to the same accuracy in $\als$ as in NLO\footnote{Note that NNLL  $\bar {\cal C}$ has the same expansion as full NNLL.} (red solid) and full NLO result (black dashed). One can see that NLO with the $qg$ channel contribution subtracted is very well (especially for $\mu_0=Q$) approximated by the expansion of resummed cross section. Such good agreement lets us conclude that the NNLL resummed formula will take into account a prevailing part of the higher-order contributions from the $q \bar q$ and $gg $ channels to all orders in $\als$.  Expansion of the resummed cross section differs significantly from the full NLO, this can be explained by the fact that contribution from the $qg$ channel appears first at NLO so no resummation is performed for this channel~\cite{Li:2014ula,Broggio:2015lya}. Even thought formally subleading w.r.t.\ contributions from the $q \bar q$ and $gg$ channels, the $qg$ production channel carries a relatively large numerical significance, especially at small scales.

\begin{table}
\begin{center}
\begin{tabular}{|c c c c c  c |}
	\hline
	$\sqrt{S}$ {[}TeV{]} & $\mu_0$ & NLO {[}fb{]} & {NLO+NLL}{[}fb{]} & {NLO+NNLL $\bar {\cal C}$} {[}fb{]} & {NLO+NNLL}{[}fb{]} \tabularnewline
	\hline 
	14  & $Q$ & $506_{-11.5\%}^{+11.8\%}$ & $530_{-9.2\%}^{+9.8\%}$  & $598_{-7.3\%}^{+7.8\%}$ &  $603_{-6.9\%}^{+7.8\%}$  \tabularnewline
	 & $Q/2$ & $566_{-10.6\%}^{+9.9\%}$ & $576_{-8.0\%}^{+8.7\%}$  & $600_{-7.0\%}^{+6.1\%}$  & $602_{-6.4\%}^{+6.0\%}$  \tabularnewline 
	& $M/2$ & $604_{-9.2\%}^{+6.1\%}$ & $609_{-7.8\%}^{+8.4\%}$  & $609_{-6.9\%}^{+6.9\%}$  & $607_{-6.1\%}^{+5.7\%}$  \tabularnewline
	\hline 
\end{tabular}
\end{center}
\caption{Total cross section predictions for $pp \to \ttbh$ at various central scale choices and resummation accuracies. The listed error is the theoretical error due to scale variation calculated using the 7-point method.}
\label{t:totalxsec}
\end{table}

\begin{figure}
\centering
\includegraphics[width=0.45\textwidth]{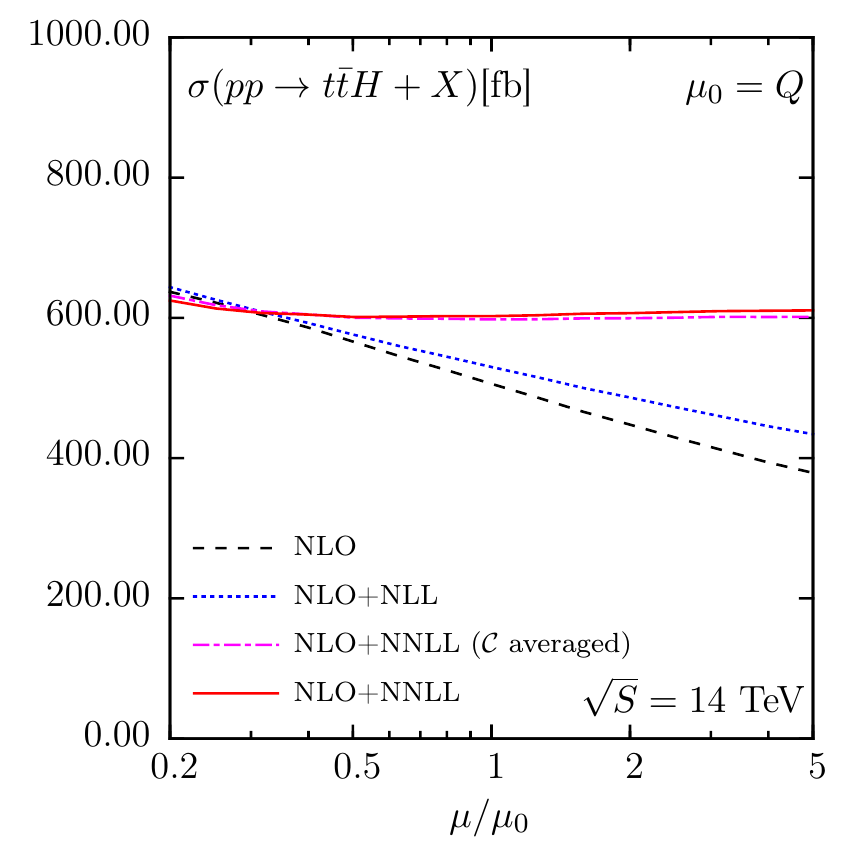}
\includegraphics[width=0.45\textwidth]{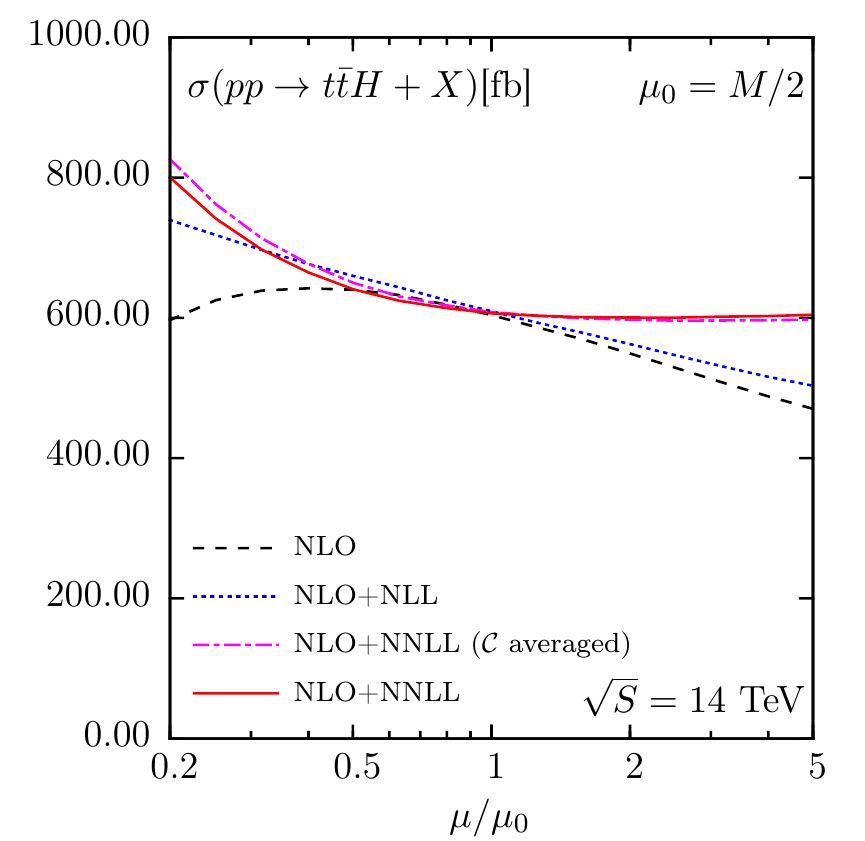}
\caption{Scale dependence of the total cross section for the process $pp\to \ttbh$ at the LHC with $\sqrt S=14$ TeV. Results shown for the choice $\mu=\muf=\mur$ and two central scale values $\mu_0=Q$ (left plot) and $\mu_0=M/2$ (right plot).} 
\label{f:scaledependence14}
\end{figure}

In Table \ref{t:totalxsec} we show our numerical predictions for the total cross sections for three scale choices: $\mu_0=Q$, $\mu_0=M/2$ and `in-between' value of $\mu_0=Q/2$. The theoretical error due to scale variation is calculated using the 7-point method\footnote{In 7-point method error is calculated from minimum and maximum values obtained with $(\muf/\mu_{0}, \mur/\mu_{0}) = (0.5,0.5), (0.5,1), (1,0.5), (1,1), (1,2), (2,1), (2,2)$.}.
It can be seen that for all scale choices the theoretical error decreases when one improves the predictions by adding resummation. For example, for $\mu_0=Q/2$ the theoretical precision of the NLO+NNLL prediction is improved by about 40\% with respect to the NLO result, bringing the scale error calculated with the 7-point method down to less than 6.5\% of the central cross section value. Comparing last two columns of Table~\ref{t:totalxsec} we can conclude that the averaging of non-logarithmic contributions and removing $\mathbf{H}^{\mathrm{(1)}}_{R} \mathbf{\tilde S}^{\mathrm{(1)}}_{R}$ term result in a difference of below~1$\%$.

In Figure~\ref{f:scaledependence14} we show the scale dependence of $\ttbh$ total cross sections calculated with the factorization and renormalization scale kept equal, $\mu=\muf=\mur$. We observe a substantial increase in the stability of the cross section value w.r.t.\ scale variation as the accuracy of resummation improves from NLL to NNLL. The NLO+NNLL prediction is characterised by a very low scale dependence. The rise of the cross section at small scales (for $\mu_0=M/2$) is driven by the fall of the expansion of resummed result NNLL$|_{\rm NLO}$  (second term in Eq.~(\ref{invmel})) and is a consequence of the relatively large scale dependence of NLO $qg$ channel contribution. This contribution appears first at NLO so no resummation is performed for it. %Even thought formally subleading w.r.t.\ contributions from the $q \bar q$ and $gg$ channels, the $qg$ production channel carries a relatively large numerical significance at small scales~\cite{Kulesza:2017ukk,Broggio:2015lya}.
Even though the $qg$ production channel is formally subleading w.r.t $q \bar q$ and $gg$ channels, it carries a relatively large numerical significance at low scales~\cite{Kulesza:2017ukk,Broggio:2015lya}. Furthermore, we see that the colour-averaging procedure introduced in Eqs. (\ref{Ccoeff_Def}) and (\ref{eq:res:fact_diag}) has only a minimal impact on the numerical results, i.e. NNLL $\bar {\cal C}$ results provide a very good approximation of the full NNLL results.

%At the end of this section we compare our results with Broggio et al.~\cite{Broggio:2016lfj}. For this purpose we calculated our full NLO+NNLL cross-section for $\sqrt S=13$ TeV with MMHT2014 pdf sets and $\mu_0=Q$.\footnote{The scale choices made in our approach and \cite{Broggio:2016lfj} are not equivalent in general, they coincide only for $\mu_0=Q$. The scale errors were calculated using different methods.}
%We obtain $\sigma_{\rm NLO+NNLL}=501.7^{+38.6}_{-34.6}$ fb, to be compared with $514.3^{+42.9}_{-39.5}$ fb reported in \cite{Broggio:2016lfj}, i.e. the central results of the two calculations agree within 2.5\%.

\section*{Acknowledgments}

We are grateful to M. Kr\"amer for providing us with a numerical code for  NLO $\ttbh$ cross section calculations~\cite{Beenakker:2001rj}. This work has been supported in part by the DFG grant KU 3103/1. Support of the Polish National Science Centre grant no.\ DEC-2014/13/B/ST2/02486 is gratefully acknowledged. TS acknowledges support in the form of a Westf\"alische Wilhelms-Universit\"at Internationalisation scholarship.
This work was also partially supported by the U.S.\ National Science Foundation, under grants PHY-0969510, the LHC Theory Initiative, PHY-1417317 and PHY-1619867. TS would like to thank the organizers of the EPS-HEP 2017 conference for the very interesting meeting and for the possibility to present this talk.

\end{document}